\documentclass[11pt]{revtex4}
\usepackage{amssymb,epsf}
\usepackage{latexsym}
\begin{document}

\title{Thermodynamical properties of topological Born-Infeld-dilaton black holes}
\author{Ahmad Sheykhi \footnote{sheykhi@mail.uk.ac.ir}}
\address{Department of Physics, Shahid Bahonar University, P.O. Box 76175-132, Kerman, Iran\\
Research Institute for Astronomy and Astrophysics of Maragha
(RIAAM), Maragha, Iran}

\begin{abstract}
We examine the $(n+1)$-dimensional $(n\geq3)$ action in which
gravity is coupled to the Born-Infeld nonlinear electrodynamic and
a dilaton field. We construct a new $(n+1)$-dimensional analytic
solution of this theory in the presence of Liouville-type dilaton
potentials. These solutions which describe charged topological
dilaton black holes with nonlinear electrodynamics, have unusual
asymptotics. They are neither asymptotically flat nor (anti)-de
Sitter. The event horizons of these black holes can be an
$(n-1)$-dimensional positive, zero or negative constant curvature
hypersurface. We also analyze thermodynamics and stability of
these solutions and disclose the effect of the dilaton and
Born-Infeld fields on the thermal stability in the canonical
ensemble.

\end{abstract}

\maketitle

\section{Introduction}
The pioneering theory of the non-linear electromagnetic field was
proposed by Born and Infeld in $1934$ for the purpose of solving
various problems of divergence appearing in the Maxwell theory
\cite{BI}. Although it became less popular with the introduction
of QED, in recent years, the Born-Infeld action has been occurring
repeatedly with the development of superstring theory, where the
dynamics of D-branes is governed by the Born-Infeld action
\cite{Frad,Cal}. It has been shown that charged black hole
solutions in Einstein-Born-Infeld gravity are less singular in
comparison with
the Reissner-Nordstr\"{o}m solution. In other words, there is no Reissner-Nordstr%
\"{o}m-type divergence term $q^{2}/r^{2}$ in the metric near the
singularity while it exist only a Schwarzschild-type term $m/r$
\cite{Car,Banad}. In the absence of a dilaton field, exact
solutions of Einstein-Born-Infeld theory with/without cosmological
constant have been constructed by many authors
\cite{Gar,Wil,Aie,Tamaki,Fern,Dey,Cai1, Deh3}. In the
scalar-tensor theories of gravity, black hole solutions coupled to
a Born-Infeld nonlinear electrodynamics have been studied in
\cite{yaz3}. The Born-Infeld action coupled to a dilaton field,
appears in the low energy limit of open superstring theory
\cite{Frad}. Although one can consistently truncate such models,
the presence of the dilaton field cannot be ignored if one
consider coupling of the gravity to other gauge fields, and
therefore one remains with Einstein-Born-Infeld gravity in the
presence of a dilaton field. Many attempts have been done to
construct solutions of Einstein-Born-Infeld-dilaton (EBId) gravity
\cite{YI,Tam,yaz,SRM,Shey,DHSR}. The appearance of the dilaton
field changes the asymptotic behavior of the solutions to be
neither asymptotically flat nor (anti)-de Sitter [(A)dS]. The
motivation for studying non-asymptotically flat nor (A)dS
solutions of Einstein gravity comes from the fact that, these kind
of solutions can shed some light on the possible extensions of
AdS/CFT correspondence. Indeed, it has been speculated that the
linear dilaton spacetimes, which arise as near-horizon limits of
dilatonic black holes, might exhibit holography \cite{Ahar}. Black
hole spacetimes which are neither asymptotically flat nor (A)dS
have been explored widely in the literature (see e.g.
\cite{CHM,Cai,Clem,Shey0, SR,DF,SDR,SDRP,yaz2}).

On the other hand, it is a general belief that in four dimensions
the topology of the event horizon of an asymptotically flat
stationary black hole is uniquely determined to be the two-sphere
$S^2$ \cite{Haw1,Haw2}. Hawking's theorem requires the integrated
Ricci scalar curvature with respect to the induced metric on the
event horizon to be positive \cite{Haw1}. This condition applied
to two-dimensional manifolds determines uniquely the topology. The
``topological censorship theorem" of Friedmann, Schleich and Witt
is another indication of the impossibility of non spherical
horizons \cite{FSW1,FSW2}. However, when the asymptotic flatness
and the four dimensional spacetime are given up, there are no
fundamental reasons to forbid the existence of static or
stationary black holes with nontrivial topologies. It was
confirmed that black holes in higher dimensions bring rich physics
in comparison with the four dimensions. For instance, for
five-dimensional asymptotically flat stationary black holes, in
addition to the known $S^3$ topology of event horizons, stationary
black hole solutions with event horizons of $S^2 \times S^1$
topology (black rings) have been constructed \cite{Emp}. It has
been shown that for asymptotically AdS spacetime, in the
four-dimensional Einstein-Maxwell theory, there exist black hole
solutions whose event horizons may have zero or negative constant
curvature and their topologies are no longer the two-sphere $S^2$.
The properties of these black holes are quite different from those
of black holes with usual spherical topology horizon, due to the
different topological structures of the event horizons. Besides,
the black hole thermodynamics is drastically affected by the
topology of the event horizon. It was argued that the Hawking-Page
phase transition \cite{Haw3} for the Schwarzschild-AdS black holes
does not occur for locally AdS black holes whose horizons have
vanishing or negative constant curvature, and they are thermally
stable \cite{Birm}. The studies on the topological black holes
have been carried out extensively in many aspects (see e.g.
\cite{Lemos,Cai2,Bril1,Cai3,Cai4,Cri,other,Ban,Sheykhi}).

In this paper, we would like to explore thermodynamical properties
of the topological Born-Infeld-dilaton black holes in higher
dimensional spacetimes in the presence of Liouville-type
potentials for the dilaton field. The motivation for studying
higher dimensional solutions of Einstein gravity originates from
superstring theory, which is a promising candidate for the unified
theory of everything. As the superstring theory can be
consistently formulated only in $10$-dimensions, the existence of
extra dimensions should be regarded as the prediction of the
theory. Although for a while it was thought that the extra spatial
dimensions would be of the order of the Planck scale, making a
geometric description unreliable, but it has recently been
realized that there is a way to make the extra dimensions
relatively large and still be unobservable. This is if we live on
a three dimensional surface (brane) in a higher dimensional
spacetime (bulk). In such a scenario, all gravitational objects
such as black holes are higher dimensional. Indeed, the large
extra dimension scenarios open up new exciting possibilities to
relate the properties of higher dimensional black holes to the
observable world by direct probing of TeV-size mini-black holes at
future high energy colliders \cite{Gid}. Besides, it was argued
that through Hawking radiation of higher dimensional black holes,
it is possible to detect these extra dimensions \cite{Ishi}. In
the light of all mentioned above, it becomes obvious that further
study of black hole solutions in higher dimensional gravity is of
great importance.

This paper is organized as follows: In section \ref{Field}, we
construct a new class of $(n+1)$-dimensional topological black
hole solutions in EBId theory with two liouville type potentials
and general dilaton coupling constant, and investigate their
properties. In section \ref{Therm}, we obtain the conserved and
thermodynamic quantities of the $(n+1)$-dimensional topological
black holes and verify that these quantities satisfy the first law
of black hole thermodynamics. In section \ref{stab}, we perform a
stability analysis and show that the dilaton creates an unstable
phase for the solutions. The last section is devoted to summary
and conclusions.

\section{Topological dilaton black holes in Born-Infeld theory}\label{Field}

We examine the $(n+1)$-dimensional $(n\geq3)$ action in which
gravity is coupled to dilaton and Born-Infeld fields
\begin{equation}\label{Act}
S=\frac{1}{16\pi}\int{d^{n+1}x\sqrt{-g}\left(\mathcal{R}\text{
}-\frac{4}{n-1}(\nabla \Phi )^{2}-V(\Phi )+L(F,\Phi)\right)},
\end{equation}
where $\mathcal{R}$ is the Ricci scalar curvature, $\Phi $ is the
dilaton field and $V(\Phi )$ is a potential for $\Phi $. The
Born-Infeld $L(F,\Phi)$ part of the action is given by
\begin{equation}
L(F,\Phi )=4\beta ^{2}e^{4\alpha \Phi /(n-1)}\left( 1-\sqrt{1+\frac{%
e^{-8\alpha \Phi /(n-1)}F^{2}}{2\beta ^{2}}}\right).
\end{equation}
Here, $\alpha $ is a constant determining the strength of coupling
of the scalar and electromagnetic fields, $F^2=F_{\mu \nu }F^{\mu
\nu }$, where $F_{\mu \nu }=\partial _{\mu }A_{\nu }-\partial
_{\nu }A_{\mu }$ is the electromagnetic field tensor, and $A_{\mu
}$ is the electromagnetic vector potential. $\beta $ is the
Born-Infeld parameter with the dimension of mass. In the limit
$\beta \rightarrow \infty $, $L(F,\Phi)$ reduces to the standard
Maxwell field coupled to a dilaton field
\begin{equation}
L(F,\Phi )=-e^{-4\alpha \Phi /(n-1)}F^{2}.
\end{equation}
On the other hand, $L(F,\Phi )\rightarrow 0$ as $\beta \rightarrow
0$. It is convenient to set
\begin{equation}
L(F,\Phi )=4\beta ^{2}e^{4\alpha \Phi /(n-1)}{\mathcal{L}}(Y),
\end{equation}
where
\begin{eqnarray}
{\mathcal{L}}(Y) &=&1-\sqrt{1+Y},  \label{LY} \\
Y &=&\frac{e^{-8\alpha \Phi /(n-1)}F^{2}}{2\beta ^{2}}.  \label{Y}
\end{eqnarray}
The equations of motion can be obtained by varying the action
(\ref{Act}) with respect to the gravitational field $g_{\mu \nu
}$, the dilaton field $\Phi $ and the gauge field $A_{\mu }$ which
yields the following field equations
\begin{eqnarray}
\mathcal{R}_{\mu \nu } &=&\frac{4}{n-1}\left( \partial _{\mu }\Phi
\partial _{\nu }\Phi +\frac{1}{4}g_{\mu \nu }V(\Phi )\right)
-4e^{-4\alpha \Phi /(n-1)}\partial _{Y}{\mathcal{L}}(Y)F_{\mu \eta
}F_{\nu }^{\text{ }\eta }
\nonumber \\
&&+\frac{4\beta ^{2}}{n-1}e^{4\alpha \Phi /(n-1)}\left[ 2Y\partial _{Y}{%
\mathcal{L}}(Y)-{\mathcal{L}}(Y)\right] g_{\mu \nu },  \label{FE1}
\end{eqnarray}
\begin{equation}
\nabla ^{2}\Phi =\frac{n-1}{8}\frac{\partial V}{\partial \Phi
}+2\alpha
\beta ^{2}e^{4\alpha \Phi /(n-1)}\left[ 2{\ Y}\partial _{Y}{\mathcal{L}}(Y)-%
\mathcal{L}(Y)\right] ,  \label{FE2}
\end{equation}
\begin{equation}
\partial _{\mu }\left( \sqrt{-g}e^{-4\alpha \Phi /(n-1)}\partial _{Y}{%
\mathcal{L}}(Y)F^{\mu \nu }\right) =0.  \label{FE3}
\end{equation}
In particular, in the case of the linear electrodynamics with
${\cal L}(Y)=-{1\over 2}Y$, the system of equations
(\ref{FE1})-(\ref{FE3}) reduce to the well-known equations of
Einstein-Maxwell-dilaton gravity \cite{CHM}.\\
We would like to find topological solutions of the above field
equations. The most general such metric can be written in the form
\begin{equation}\label{metric}
ds^2=-f(r)dt^2 + {dr^2\over f(r)}+ r^2R^2(r)h_{ij}dx^{i}dx^{j} ,
\end{equation}
where $f(r)$ and $R(r)$ are functions of $r$ which should be
determined, and $h_{ij}$ is a function of coordinates $x_{i}$
which spanned an $(n-1)$-dimensional hypersurface with constant
scalar curvature $(n-1)(n-2)k$. Here $k$ is a constant and
characterizes the hypersurface. Without loss of generality, one
can take $k=0, 1, -1$, such that the black hole horizon or
cosmological horizon in (\ref{metric}) can be a zero (flat),
positive (elliptic) or negative (hyperbolic) constant curvature
hypersurface. The electromagnetic field equation (\ref{FE3}) can
be integrated immediately to give
\begin{equation}\label{Ftr}
F_{tr}=\frac{\beta q e^{4\alpha \Phi/(n-1)}}{\sqrt{\beta^2 \left(
rR\right)^{2n-2}+q^{2}}},
\end{equation}
where $q$ is an integration constant related to the electric
charge of the black hole. Defining the electric charge via $ Q =
\frac{1}{4\pi} \int \exp\left[{-4\alpha\Phi/(n-1)}\right]  \text{
}^{*} F d{\Omega},$ we get
\begin{equation}
{Q}=\frac{q\omega _{n-1}}{4\pi},  \label{Charge}
\end{equation}
where $\omega_{n-1}$ represents the volume of constant curvature
hypersurface described by $h_{ij}dx^idx^j$ . It is worthwhile to
note that the electric field is finite at $r=0$. This is expected
in Born-Infeld theories. Meanwhile it is interesting to consider
three limits of (\ref{Ftr}). First, for large $\beta$ (where the
Born-Infeld action reduces to the Maxwell case) we have
$F_{tr}=\frac{q e^{4\alpha \Phi/(n-1)}}{(rR)^{n-1}}$ as presented
in \cite{CHM}. On the other hand, if $\beta\rightarrow 0$ we get
$F_{tr}=0$. Finally, in the absence of the dilaton field
($\alpha=0$), it reduces to the case of $(n+1)$-dimensional
Einstein-Born-Infeld theory \cite{Dey}
\begin{equation}
F_{tr}=\frac{ \beta q }{\sqrt{\beta^2 r^{2n-2}+q^{2}}}.
\end{equation}
Our aim here is to construct exact, $(n+1)$-dimensional
topological solutions of the EBId gravity with an arbitrary
dilaton coupling parameter $\alpha$. The case in which we find
topological solutions of physical interest is to take the dilaton
potential of the form
\begin{equation}\label{v2}
V(\Phi) = 2\Lambda_{0} e^{2\zeta_{0}\Phi} +2 \Lambda e^{2\zeta
\Phi},
\end{equation}
where $\Lambda_{0}$,  $\Lambda$, $ \zeta_{0}$ and $ \zeta$ are
constants. This kind of  potential was previously investigated by
a number of authors both in the context of
Friedmann-Robertson-Walker (FRW) scalar field cosmologies
\cite{ozer} and dilaton black holes (see e.g.
\cite{CHM,yaz2,SRM,Sheykhi,Shey}). In order to solve the system of
equations (\ref{FE1}) and (\ref{FE2}) for three unknown functions
$f(r)$, $R(r)$ and $\Phi (r)$, we make the ansatz
\begin{equation}
R(r)=e^{2\alpha \Phi /(n-1)}.\label{Rphi}
\end{equation}
Using (\ref{Rphi}), the electromagnetic field (\ref{Ftr}) and the
metric (\ref{metric}), one can show that equations (\ref{FE1}) and
(\ref{FE2}) have solutions of the form
\begin{eqnarray}
f(r)&=&-{\frac { k(n-2)\left( { \alpha}^{2}+1 \right)
^{2}{b}^{-2\gamma}}{\left( {
\alpha}^{2}-1 \right)  \left(n+{\alpha}^{2}-2 \right) }}{r}^{2\gamma}-\frac{m}{%
r^{(n-1)(1-\gamma )-1}}+\frac{2\left( \Lambda -2\beta ^{2}\right)
(\alpha ^{2}+1)^{2}b^{2\gamma }}{(n-1)(\alpha
^{2}-n)}r^{2-2\gamma}\nonumber \\
&&-\frac{4\beta^2(\alpha ^{2}+1)b^{2\gamma }}{n-1}r^{(n-1)(\gamma
-1)+1}\int r ^{(n+1)(1-\gamma)-2}\sqrt{1+\eta}{dr },  \label{f}
\end{eqnarray}
\begin{equation}
\Phi (r)=\frac{(n-1)\alpha }{2(1+\alpha ^{2})}\ln (\frac{b}{r}),
\label{phi}
\end{equation}
where $b$ is an arbitrary constant, $\gamma =\alpha ^{2}/(\alpha
^{2}+1)$, and
\begin{equation}
\eta = \frac{q^{2}b^{2\gamma (1-n)}}{\beta ^{2}r^{2(n-1)(1-\gamma
)}}. \label{eta}
\end{equation}
In the above expression, $m$ appears as an integration constant
and is related to the ADM (Arnowitt-Deser-Misner) mass of the
black hole. According to the definition of mass due to Abbott and
Deser \cite{abot}, the mass of the solution (\ref{f}) is
\cite{Sheykhi}
\begin{equation}
{M}=\frac{b^{(n-1)\gamma}(n-1) \omega _{n-1}}{16\pi(\alpha^2+1)}m.
\label{Mass}
\end{equation}
In order to fully satisfy the system of equations, we must have
\begin{equation}\label{lam}
\zeta_{0} =\frac{2}{\alpha(n-1)},   \hspace{.8cm}
\zeta=\frac{2\alpha}{n-1}, \hspace{.8cm}    \Lambda_{0} =
\frac{k(n-1)(n-2)\alpha^2 }{2b^2(\alpha^2-1)}.
\end{equation}
Notice that here  $\Lambda$ is a free parameter which plays the
role of the cosmological constant. For later convenience, we
redefine it as $\Lambda=-n(n-1)/2l^2$, where $l$ is a constant
with dimension of length. The integral can be done in terms of
hypergeometric function and can be written in a compact form as
\begin{eqnarray}\label{f2}
f(r) &=&-{\frac { k\left(n-2 \right)\left( { \alpha}^{2}+1 \right)
^{2}{b}^{-2\gamma}}{\left( { \alpha}^{2}-1 \right)
\left(n+{\alpha}^{2}-2 \right)
}}{r}^{2\gamma}-\frac{m}{r^{(n-1)(1-\gamma )-1}}+\frac{2\Lambda
\left( {\alpha}^{2}+1 \right) ^{2}{b}^{2
\gamma}}{(n-1)(\alpha^{2}-n )}r^{2(1-\gamma)} \nonumber
\\
&&-\frac{4\beta^2 (\alpha ^{2}+1)^{2}b^{2\gamma }r^{2(1-\gamma
)}}{(n-1)(\alpha ^{2}-n)}\times
\left( 1-\text{{\ }}_{2}F_{1}\left( \left[ -\frac{1}{2},\frac{\alpha ^{2}-n%
}{2n-2}\right] ,\left[ \frac{\alpha ^{2}+n-2}{2n-2}\right]
,-\eta\right) \right).
\end{eqnarray}
One may note that as $\beta \longrightarrow \infty $ these
solutions reduce to the $(n+1)$-dimensional topological dilaton
black hole solutions given in  \cite{Sheykhi}
\begin{eqnarray}
f(r)&=&-{\frac { k(n-2)\left( { \alpha}^{2}+1 \right)
^{2}{b}^{-2\gamma}}{\left( { \alpha}^{2}-1 \right)
\left({\alpha}^{2}+n-2 \right)
}}{r}^{2\gamma}-\frac{m}{r^{(n-1)(1-\gamma
)-1}}+\frac{2\Lambda (\alpha ^{2}+1)^{2}b^{2\gamma }}{(n-1)(\alpha ^{2}-n)}%
r^{2(1-\gamma )}\nonumber\\
&&+\frac{2q^{2}(\alpha ^{2}+1)^{2}b^{-2(n-2)\gamma }}{(n-1)(\alpha
^{2}+n-2)}r^{2(n-2)(\gamma -1)}. \label{fEMd}
\end{eqnarray}
In the absence of a nontrivial dilaton ($\alpha=\gamma=0$), the
solution (\ref{f2}) reduces to
\begin{eqnarray}
f(r) &=&k-\frac{m}{r^{n-2}}+\frac{r^2}{l^2}+\frac{4\beta^2
r^2}{n(n-1)}\times
\left( 1- \text{{\ }}_{2}F_{1}\left( \left[ -\frac{1}{2},\frac{n%
}{2-2n}\right] ,\left[ \frac{n-2}{2n-2}\right] ,-\frac{q^{2}}{
\beta^2 r^{2n-2}}\right)\right)
\end{eqnarray}
which describes an $(n+1)$-dimensional asymptotically (A)dS
topological Born-Infeld black hole with a positive, zero or
negative constant curvature hypersurface \cite{Cai1}. Using the
fact that $_2F_1(a,b,c,z)$ has a convergent series expansion for
$|z| <1$, we can find the behavior of the metric for large $r$.
This is given by
\begin{eqnarray}
f(r) &=&-{\frac { k\left(n-2 \right)\left({ \alpha}^{2}+1 \right)
^{2}{b}^{-2\gamma}}{\left( { \alpha}^{2}-1 \right)
\left(n+{\alpha}^{2}-2 \right)
}}{r}^{2\gamma}-\frac{m}{r^{(n-1)(1-\gamma )-1}}+\frac{2\Lambda
\left( {\alpha}^{2}+1 \right) ^{2}{b}^{2
\gamma}}{(n-1)(\alpha^{2}-n )}r^{2(1-\gamma)} \nonumber \\
&& +\frac{2q^{2}(\alpha^2+1)^{2}b^{-2(n-2)\gamma}}{(n-1)(\alpha
^{2}+n-2)r^{2(n-2)(1-\gamma )}} - \frac{q^{4}(\alpha^{2}+1
)^{2}b^{-2(2n-3)\gamma }}{2\beta^2 (n-1)(\alpha
^{2}+3n-4)r^{2(2n-3)(1-\gamma )}}.
\end{eqnarray}
The last term in the right hand side of the above expression is
the leading Born-Infeld correction to the topological black hole
with dilaton field \cite{Sheykhi}. Note that for $\alpha =\gamma=
0$, the above expression reduces to
\begin{eqnarray}
f(r)
&=&k-\frac{m}{r^{n-2}}+\frac{r^2}{l^2}+\frac{2q^{2}}{(n-1)(n-2)r^{2(n-2)}}
- \frac{q^{4}}{2\beta^2 (n-1)(3n-4)r^{2(2n-3)}},
\end{eqnarray}
which has the form of the $(n+1)$-dimensional topological black
hole in  (A)dS spacetime in the limit $\beta\rightarrow \infty$
(see e.g. \cite{Bril1,Cai3}).

\begin{figure}[tbp]
\epsfxsize=7cm \centerline{\epsffile{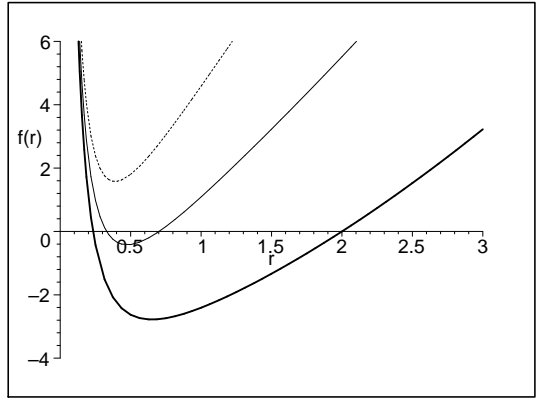}} \caption{The
function $f(r)$ versus $r$ for $\protect\alpha=0.7$, $m=2$,
$\beta=1$, $n=4$ and $q=1$. $k=-1$ (bold line), $k=0$ (continuous
line), and $k=1$ (dashed line).} \label{figure1}
\end{figure}

\begin{figure}[tbp]
\epsfxsize=7cm \centerline{\epsffile{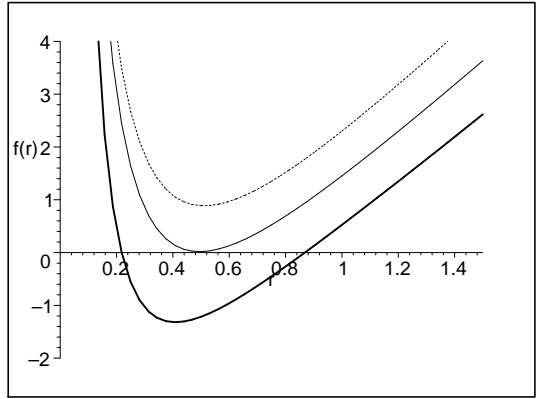}} \caption{The
function $f(r)$ versus $r$ for $m=2$, $\beta=1$, $q=1$, $n=4$ and
$k=0$. $\protect\alpha=0.6$ (bold line), $\protect\alpha=0.75$
(continuous line), and $\protect\alpha=0.85$ (dashed line).}
\label{figure2}
\end{figure}

\begin{figure}[tbp]
\epsfxsize=7cm \centerline{\epsffile{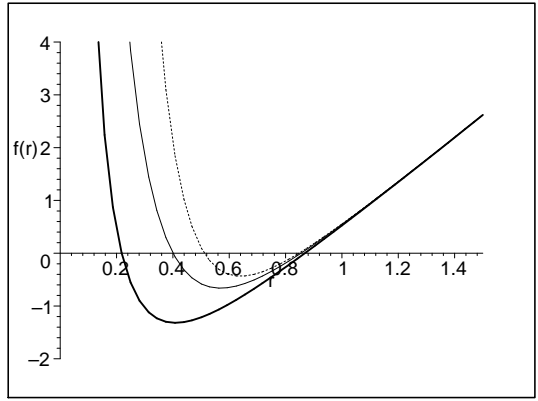}} \caption{The
function $f(r)$ versus $r$ for $m=2$, $\protect\alpha=0.6$ ,
$q=1$, $n=4$ and $k=0$. $\beta=1$ (bold line), $\beta=2$
(continuous line), and $\beta=15$ (dashed line).} \label{figure3}
\end{figure}
\begin{figure}[tbp]
\epsfxsize=7cm \centerline{\epsffile{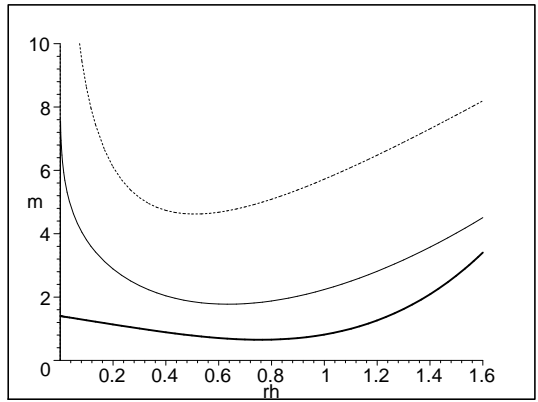}} \caption{The
function $m(r_h)$ versus $r_h$ for  $\beta=1$, $q=1$, $n=4$ and
$k=0$. $\protect\alpha=0$ (bold line), $\protect\alpha=0.8$
(continuous line), and $\protect\alpha=1.2$ (dashed line).}
\label{figure4}
\end{figure}

\begin{figure}[tbp]
\epsfxsize=7cm \centerline{\epsffile{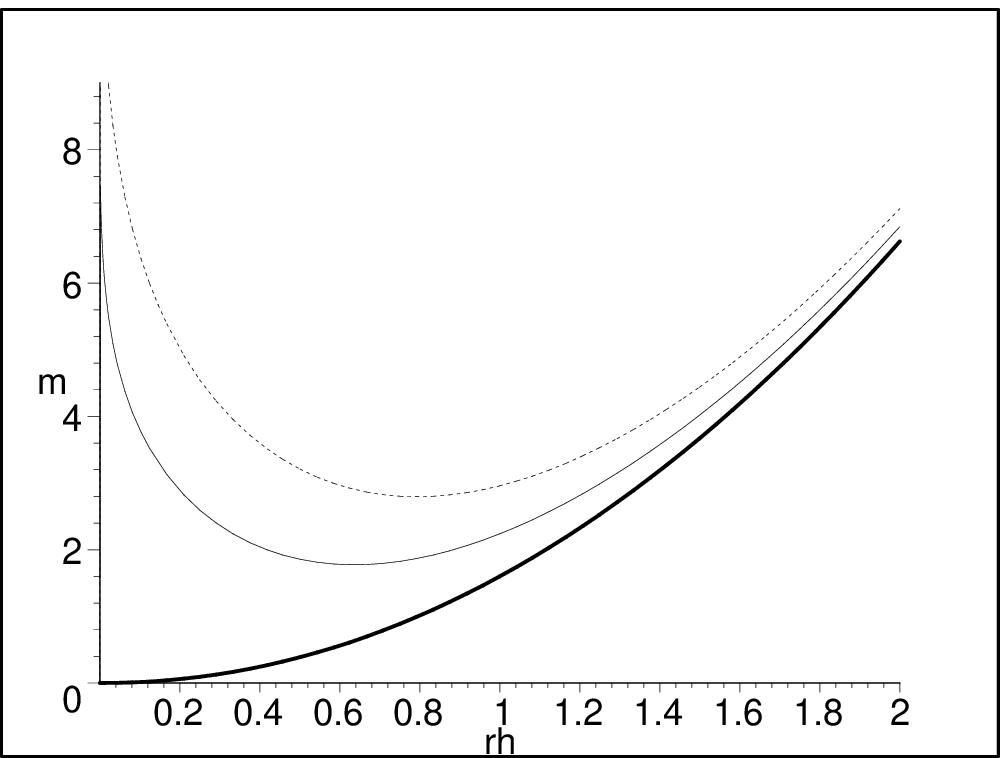}} \caption{The
function $m(r_h)$ versus $r_h$ for  $\beta=1$,
$\protect\alpha=0.8$, $n=4$ and $k=0$. $q=0$  (bold line), $q=1$
(continuous line), and $q=1.5$  (dashed line).} \label{figure5}
\end{figure}

\section*{Physical Properties of the Solutions}
In order to study the physical properties of the solutions, we
first look for the curvature singularities. In the presence of a
dilaton field, the Kretschmann scalar $R_{\mu \nu \lambda \kappa
}R^{\mu \nu \lambda \kappa }$ diverges at $r=0$, it is finite for
$r\neq 0$ and goes to zero as $r\rightarrow \infty $. Thus, there
is an essential singularity located at $r=0$. The spacetime is
neither asymptotically flat nor (A)dS. It is notable to mention
that in the $k=\pm1$ cases these solutions do not exist for the
string case where $\alpha=1$. As one can see from Eq. (\ref{f2}),
the solution is ill-defined for $\alpha =\sqrt{n}$. The cases with
$\alpha <\sqrt{n}$ and $\alpha
>\sqrt{n}$ should be considered separately. In the first case
where $ \alpha <\sqrt{n}$,  there exist a cosmological horizon for
$\Lambda >0$, while there is no cosmological horizons if $\Lambda
<0$. Indeed, where $\alpha <\sqrt{n}$ and $\Lambda <0$, the
spacetimes associated with the solution (\ref{f2}) exhibit a
variety of possible casual structures depending on the values of
the metric parameters (see Figs. \ref{figure1}-\ref{figure3}). For
simplicity in these figures, we kept fixed $l=b=1$. These figures
show that our solutions can represent topological black hole, with
two event horizons, an extreme topological black hole or a naked
singularity provided the parameters of the solutions are chosen
suitably. In the second case where $\alpha
>\sqrt{n}$, the spacetime has a cosmological horizon. One
can obtain the casual structure by finding the roots of $ f(r)=0$.
Unfortunately, because of the nature of the exponent in
(\ref{f2}), it is not possible to find analytically the location
of the horizons. To have further understanding on the nature of
the horizons, as an example for $k=0$, we plot in Figs.
\ref{figure4} and \ref{figure5} the mass parameter $m$ as a
function of the horizon radius for different values of dilaton
coupling constant $\alpha$ and charge parameter $q$. Again, we
have fixed $l=b=1$, for simplicity. It is easy to show that the
mass parameter of the black hole can be expressed in terms of the
horizon radius $r_h$ as
\begin{eqnarray}\label{mass}
m(r_{h}) &=&-{\frac { k\left(n-2 \right)\left( { \alpha}^{2}+1
\right) ^{2}{b}^{-2\gamma}}{\left( { \alpha}^{2}-1 \right)
\left(n+{\alpha}^{2}-2 \right)
}}{r_{h}}^{n-2+\gamma(3-n)}+\frac{2\Lambda \left( {\alpha}^{2}+1
\right) ^{2}{b}^{2 \gamma}}{(n-1)(\alpha^{2}-n
)}r_{h}^{n(1-\gamma)-\gamma}\nonumber\\
&&-\frac{4\beta^2 (\alpha ^{2}+1)^{2}b^{2\gamma }}{(n-1)(\alpha
^{2}-n)}r_{h}^{n(1-\gamma)-\gamma}\times
\left(1-\text{}_{2}F_{1}\left( \left[ -\frac{1}{2},\frac{\alpha ^{2}-n%
}{2n-2}\right] ,\left[ \frac{\alpha ^{2}+n-2}{2n-2}\right]
,-\eta\right) \right).
\end{eqnarray}
These figures show that for a given value of $\alpha$, the number
of horizons depend on the choice of the value of the mass
parameter $m$. We see that, up to a certain value of the mass
parameter $m$, there are two horizons, and as we decrease the $m$
further, the two horizons meet. In this case we get an extremal
black hole with mass $m_{\mathrm{ext}}$ (see the next section).
Figure \ref{figure4} shows that with increasing $\alpha$, the
$m_{\mathrm{ext}}$ also increases. It is worth noting that in the
limit $r_{h}\rightarrow0$ we have a nonzero value for the mass
parameter $m$. This is in contrast to the Schwarzschild black
holes in which the mass parameter goes to zero as
$r_{h}\rightarrow0$. As we have shown in figure \ref{figure5},
this is due to the effect of the charge parameter $q$ and the
nature of the Born-Infeld field, and in the case $q=0$, the mass
parameter $m$ goes to zero as $r_{h}\rightarrow0$. In summary, the
metric of Eqs. (\ref{metric}) and (\ref{f2}) can represent a
charged topological dilaton black hole with inner and outer event
horizons located at $r_{-}$ and $r_{+}$, provided
$m>m_{\mathrm{ext}}$, an extreme topological black hole in the
case of $m=m_{\mathrm{ext}}$, and a naked singularity if
$m<m_{\mathrm{ext}}$.

\section{Thermodynamics of topological dilaton black hole} \label{Therm}
We now would like to study the thermodynamical properties of the
solutions we have just found. The temperature of the black hole
can be obtained by continuing the metric to its Euclidean sector
via $t = -i\tau$ and requiring the absence of conical singularity
at the horizon. It is a matter of calculation to show that
\begin{equation}
T_{+}=\frac{\kappa}{2\pi}= \frac{f^{\text{ }^{\prime
}}(r_{+})}{4\pi},
\end{equation}
where $\kappa$ is the surface gravity. The temperature is then
\begin{eqnarray}\label{Tem}
T_{+}&=&-\frac{(\alpha ^2+1)b^{2\gamma}r_{+}^{1-2\gamma}}{2\pi
(n-1)}\left(
\frac{k(n-2)(n-1)b^{-4\gamma}}{2(\alpha^2-1)}r_{+}^{4\gamma-2}
+\Lambda -2\beta^2(1-\sqrt{1+\eta_{+}})\right)\nonumber\\
&=&-\frac{k(n-2)(\alpha ^2+1)b^{-2\gamma}}{2\pi(\alpha
^2+n-2)}r_{+}^{2\gamma-1}+ \frac{(n-\alpha ^{2})m}{4\pi(\alpha
^{2}+1)}{r_{+}}^{(n-1)(\gamma -1)}-\frac{q^2(\alpha
^{2}+1)b^{2(2-n)\gamma}}{\pi(\alpha
^2+n-2)}r_{+}^{2(2-n)(1-\gamma)-1}\nonumber\\
&&\times \text{ }_{2}F_{1}\left( %
\left[ {\frac{1}{2},\frac{{n+\alpha }^{2}{-2}}{{2n-2}}}\right] ,\left[ {%
\frac{{3n+\alpha }^{2}{-4}}{{2n-2}}}\right] ,-\eta_{+}\right),
\end{eqnarray}
where $\eta_{+}=\eta(r=r_{+})$. There is also an extreme value for
the mass parameter in which the temperature of the event horizon
of black hole is zero. It is a matter of calculation to show that
\begin{eqnarray}\label{mext}
m_{\mathrm{ext}}&=&\frac{2k(n-2)(\alpha
^2+1)^2b^{-2\gamma}}{(n-\alpha ^{2})(\alpha
^2+n-2)}r_{+}^{(2-n)(\gamma-1)+\gamma} +\frac{4q^2(\alpha
^{2}+1)^2b^{2(2-n)\gamma}}{(n-\alpha ^{2})(\alpha
^2+n-2)}r_{+}^{(3-n)(1-\gamma)-1}\nonumber\\
&&\times \text{ }_{2}F_{1}\left( %
\left[ {\frac{1}{2},\frac{{n+\alpha }^{2}{-2}}{{2n-2}}}\right] ,\left[ {%
\frac{{3n+\alpha }^{2}{-4}}{{2n-2}}}\right] ,-\eta_{+}\right).
\end{eqnarray} The entropy of
the topological black hole typically satisfies the so called area
law of the entropy which states that the entropy of the black hole
is a quarter of the event horizon area \cite{Beck}. This near
universal law applies to almost all kinds of black holes,
including dilaton black holes, in Einstein gravity \cite{hunt}. It
is a matter of calculation to show that the entropy of the
topological black hole is
\begin{equation}
{S}=\frac{b^{(n-1)\gamma}\omega _{n-1}r_{+}^{(n-1)(1-\gamma
)}}{4}.\label{Entropy}
\end{equation}
The electric potential $U$, measured at infinity with respect to
the horizon, is defined by
\begin{equation}
U=A_{\mu }\chi ^{\mu }\left| _{r\rightarrow \infty }-A_{\mu }\chi
^{\mu }\right| _{r=r_{+}},  \label{Pot}
\end{equation}
where $\chi=\partial_{t}$ is the null generator of the horizon.
One can easily show that the gauge potential $A_{t }$
corresponding to the electromagnetic field (\ref{Ftr}) can be
written as
\begin{eqnarray}\label{vectorpot}
A_{t}&=&\frac{qb^{(3-n)\gamma }}{\Upsilon r^{\Upsilon }}\text{ }_{2}F_{1}\left( %
\left[ {\frac{1}{2},\frac{{\alpha }^{2}+n-2}{{2n-2}}}\right] ,\left[ {%
\frac{{\alpha }^{2}+3n-4}{{2n-2}}}\right] ,-\eta\right),
\end{eqnarray}
where $\Upsilon =(n-3)(1-\gamma )+1$. Therefore, the electric
potential may be obtained as
\begin{equation}
U=\frac{qb^{(3-n)\gamma }}{ \Upsilon{r_{+}}^{\Upsilon }}\text{ }%
_{2}F_{1}\left( \left[ {\frac{1}{2},\frac{{\alpha}^{2}+n-2}{{2n-2}}}%
\right] ,\left[ {\frac{{\alpha }^{2}+3n-4}{{2n-2}}}\right]
,-\eta_{+}\right). \label{Pot}
\end{equation}
\begin{figure}[tbp]
\epsfxsize=7cm \centerline{\epsffile{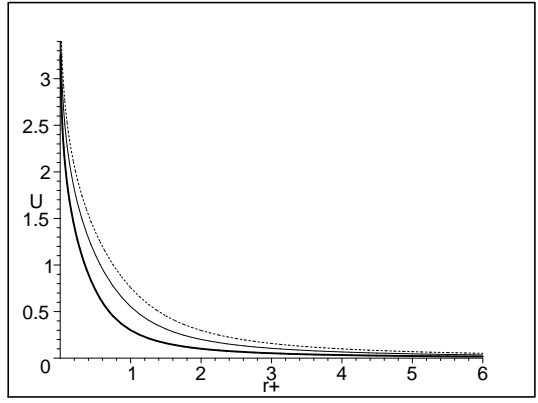}} \caption{The
function $U(r_+)$ versus $r_+$ for $\beta=1$, $n=4$, $b=1$ and
$\protect\alpha=0.8$.  $q=0.5$  (bold line), $q=1$ (continuous
line), and $q=1.5$  (dashed line).} \label{figure6}
\end{figure}
\begin{figure}[tbp]
\epsfxsize=7cm \centerline{\epsffile{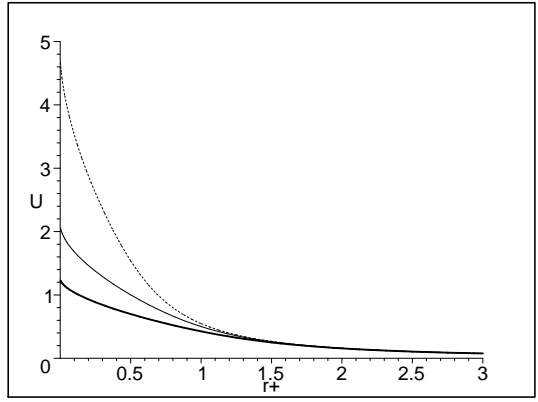}} \caption{The
function $U(r_+)$ versus $r_+$ for $q=1$, $n=4$, $b=1$ and
$\protect\alpha=0.5$.  $\beta=0.5$  (bold line), $\beta=1$
(continuous line), and $\beta=3$  (dashed line).} \label{figure7}
\end{figure}
In figures \ref{figure6} and \ref{figure7} we have shown the
behavior of the electric potential $U$ as a function of horizon
radius. As one can see from these figures, $U$ is finite even for
$r_{+}=0$. Then, we consider the first law of thermodynamics for
the topological black hole. For this purpose, we first obtain the
mass $M$ as a function of extensive quantities $S$ and $Q$. Using
the expression for the charge, the mass and the entropy given in
Eqs. (\ref{Charge}), (\ref{Mass}) and (\ref{Entropy}) and the fact
that $f(r_{+})=0$, one can obtain a Smarr-type formula as
\begin{eqnarray}
M(S,Q)&=&-\frac{k(n-1)(n-2)(\alpha^2+1)b^{-\alpha^2}
}{16\pi(\alpha^2-1)(\alpha^2+n-2)}{\left(4S\right)}^{(\alpha^2+n-2)/(n-1)}
+\frac{\Lambda(\alpha^2+1)b^{\alpha^2}}{8\pi(\alpha^2-n)}
{\left(4S\right)}^{(n-\alpha^2)/(n-1)}
\nonumber\\
&&-\frac{\beta^2(\alpha^2+1)b^{\alpha^2}}
{4\pi(\alpha^2-n)}{\left(4S\right)}^{(n-\alpha^2)/(n-1)}\times\left(1-\text{ }_{2}F_{1}\left( \left[ -\frac{1}{2},\frac{\alpha ^{2}-n%
}{2n-2}\right] ,\left[ \frac{\alpha ^{2}+n-2}{2n-2}\right] ,\frac{-\pi^2Q^{2}}{%
\beta^2 S^2}\right)\right)\nonumber \\ \label{Msmar}
\end{eqnarray}
Then we can regard the parameters $S$ and $Q$ as a complete set of
extensive parameters for the mass $M(S,Q)$ and define the
intensive parameters conjugate to $S$ and $Q$. These quantities
are the temperature and the electric potential
\begin{equation}
T=\left( \frac{\partial M}{\partial S}\right) _{Q},\ \ \  \   U=\left( \frac{\partial M%
}{\partial Q}\right) _{S}.  \label{Dsmar}
\end{equation}
Now that we have all the relevant thermodynamic quantities, we can
easily verify the first law of black hole thermodynamics. We find
that
\begin{equation}
dM = TdS+Ud{Q},
\end{equation}
is satisfied. In the next section we will explore the thermal
stability of the solutions.

\section{Stability in the canonical ensemble}\label{stab}
\begin{figure}[tbp]
\epsfxsize=7cm \centerline{\epsffile{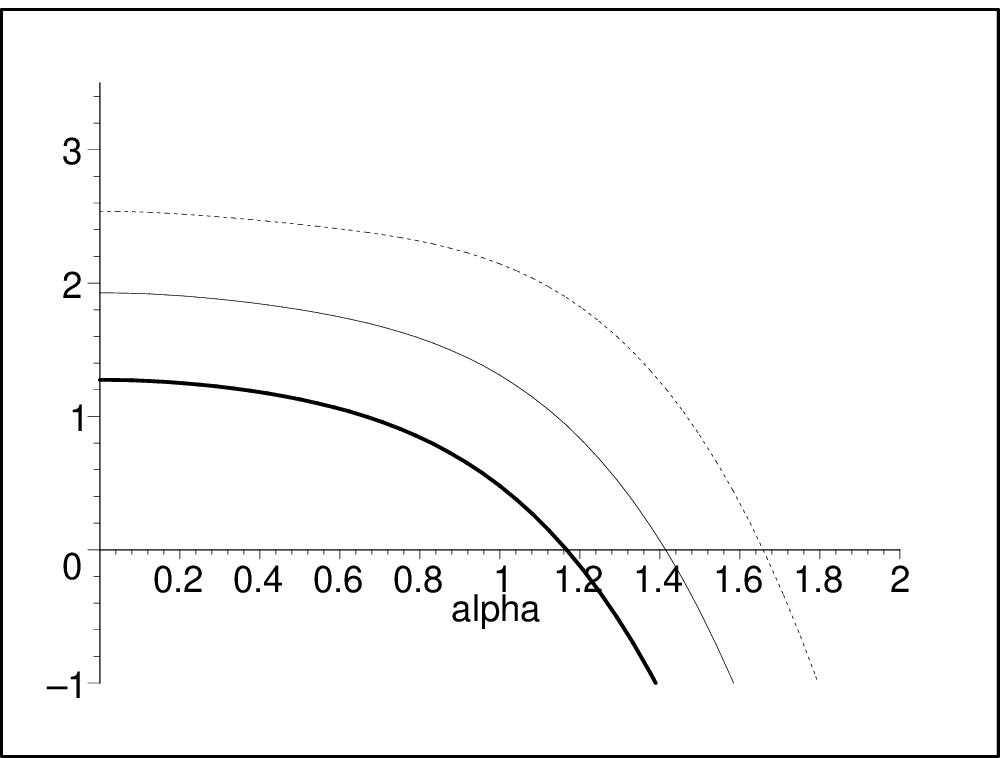}}
\caption{$(\partial ^{2}M/\partial S^{2})_{Q}$ versus
$\protect\alpha $ for $r_{+}=0.8$, $\beta=1$, $n=5$, and $k=0$.
$q=0.5$ (bold line), $q=1$ (continuous line), and $q=1.5$ (dashed
line).} \label{Figure8}
\end{figure}

\begin{figure}[tbp]
\epsfxsize=7cm \centerline{\epsffile{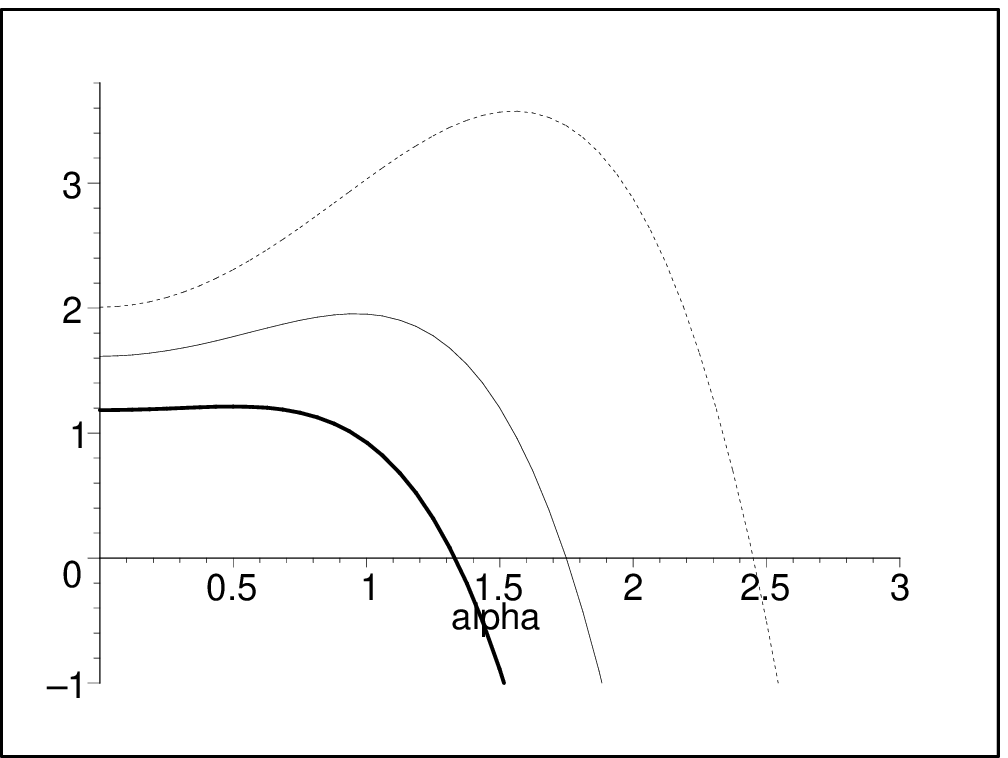}}
\caption{$(\partial ^{2}M/\partial S^{2})_{Q}$ versus
$\protect\alpha $ for $r_{+}=0.8$, $\beta=1$, $n=3$, and $k=-1$.
$q=0.5$ (bold line), $q=1$ (continuous line), and $q=1.5$ (dashed
line).} \label{Figure9}
\end{figure}
\begin{figure}[tbp]
\epsfxsize=7cm \centerline{\epsffile{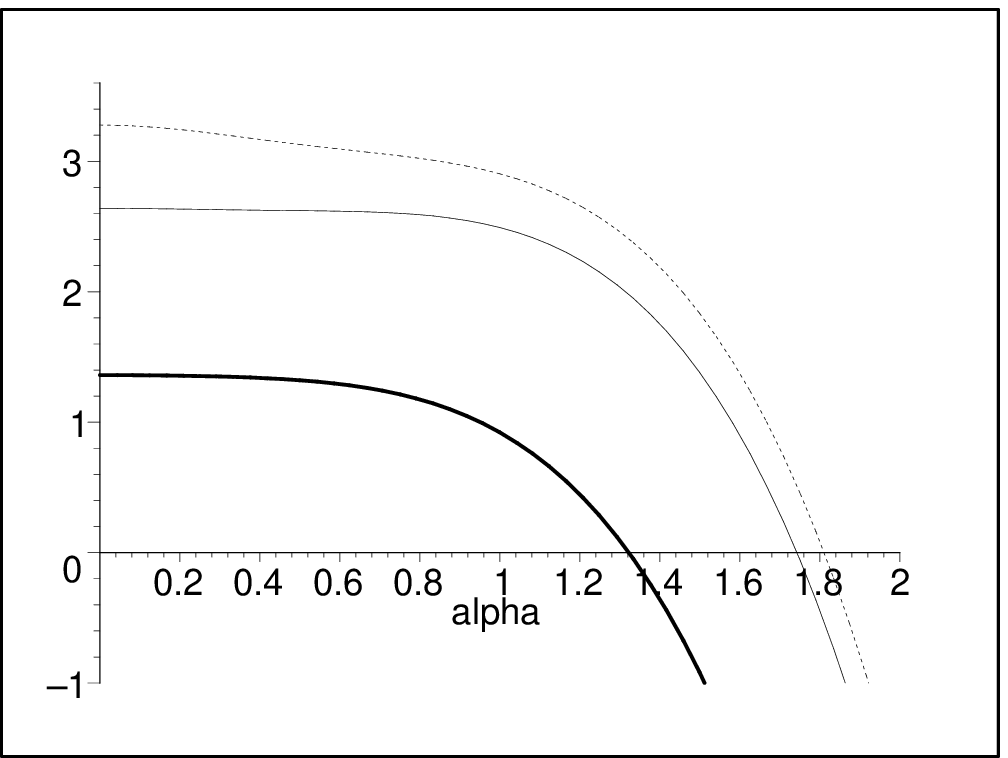}}
\caption{$(\partial ^{2}M/\partial S^{2})_{Q}$ versus
$\protect\alpha$ for $r_{+}=0.8$, $n=4$, $q=1$ and $k=-1$.
$\protect\beta=0.2$ (bold line), $\protect\beta=2$ (continuous
line) and $\protect\beta=20$ (dashed line).} \label{Figure10}
\end{figure}

\begin{figure}[tbp]
\epsfxsize=7cm \centerline{\epsffile{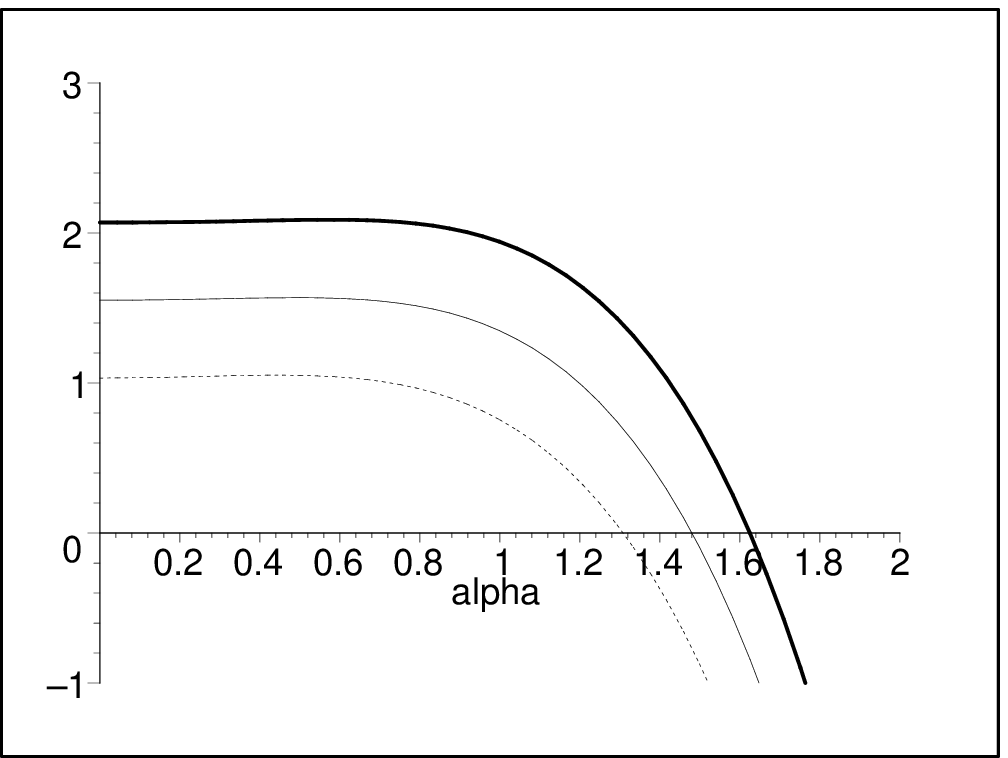}}
\caption{$(\partial ^{2}M/\partial S^{2})_{Q}$ versus
$\protect\alpha $ for $r_{+}=0.8$, $\beta=1$, $q=1$ and $n=4$.
$k=-1$ (bold line), $k=0$ (continuous line), and $k=1$ (dashed
line).} \label{Figure11}
\end{figure}

\begin{figure}[tbp]
\epsfxsize=7cm \centerline{\epsffile{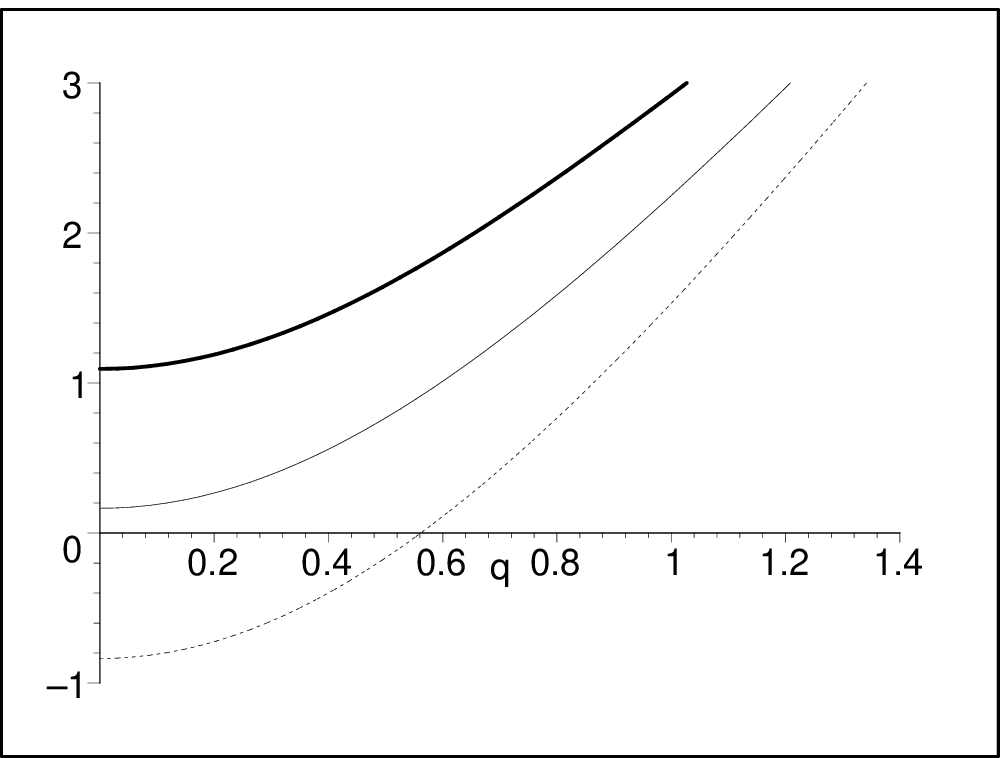}}
\caption{$(\partial ^{2}M/\partial S^{2})_{Q}$ versus $q $ for
$r_{+}=0.8$, $\beta=2$, $n=5$ and $k=-1$. $\protect\alpha=0.8$
(bold line), $\protect\alpha=1.2$ (continuous line), and
$\protect\alpha=\sqrt{2}$ (dashed line).} \label{Figure12}
\end{figure}

\begin{figure}[tbp]
\epsfxsize=7cm \centerline{\epsffile{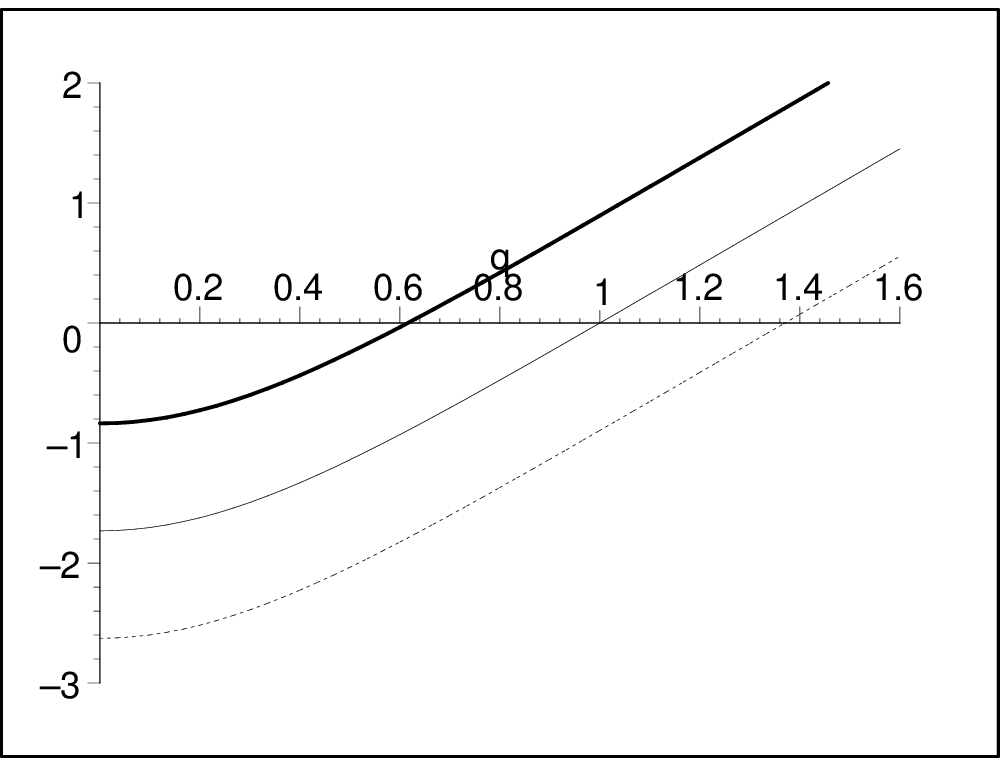}}
\caption{$(\partial ^{2}M/\partial S^{2})_{Q}$ versus $q $ for
$r_{+}=0.8$, $\beta=1$, $n=5$ and $\protect\alpha=\sqrt{2}$. $k
=-1$ (bold line), $k =0$ (continuous line), and  $k =1$ (dashed
line).} \label{Figure13}
\end{figure}

\begin{figure}[tbp]
\epsfxsize=7cm \centerline{\epsffile{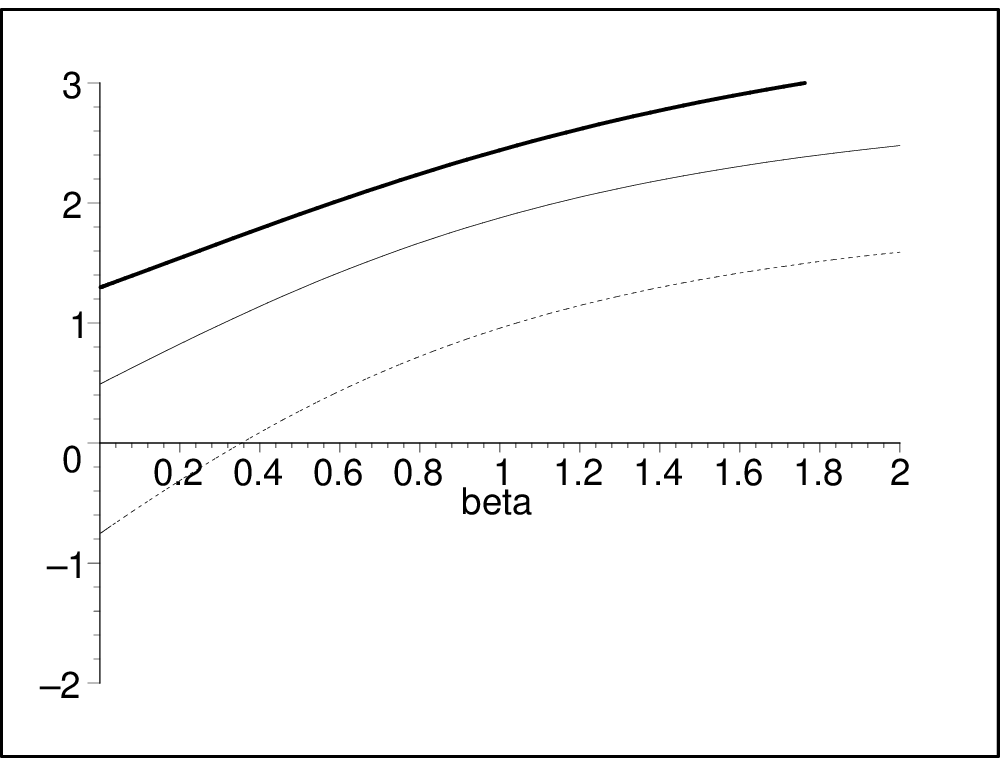}}
\caption{$(\partial ^{2}M/\partial S^{2})_{Q}$ versus $\beta $ for
$r_{+}=0.8$, $q=1$, $n=5$ and $k=-1$. $\protect\alpha=0.6$ (bold
line), $\protect\alpha=1.1$ (continuous line), and
$\protect\alpha=1.4$ (dashed line).} \label{Figure14}
\end{figure}

\begin{figure}[tbp]
\epsfxsize=7cm \centerline{\epsffile{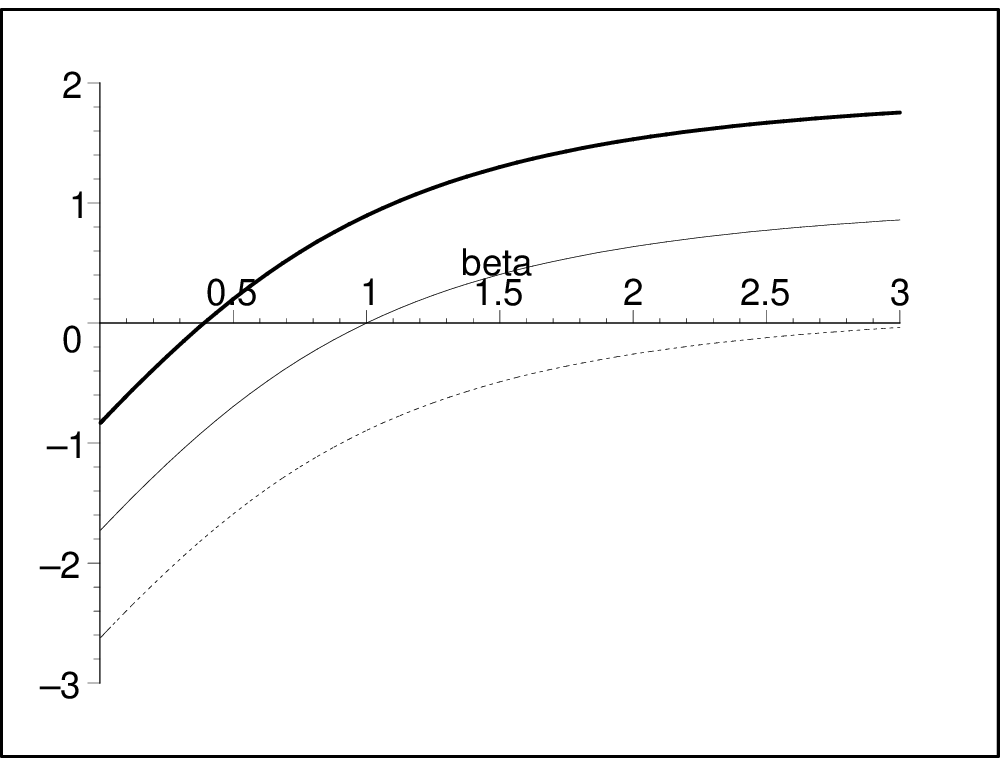}}
\caption{$(\partial ^{2}M/\partial S^{2})_{Q}$ versus $\beta $ for
$r_{+}=0.8$, $q=1$,  $\protect\alpha=\sqrt{2}$ and  $n=5$. $k=-1$
(bold line), $k=0$ (continuous line), and $k=1$ (dashed line).}
\label{Figure15}
\end{figure}
Finally, we investigate the thermal stability of the topological
black hole solutions of Einstein-Born-Infeld-dilaton gravity in
the canonical ensemble. In the canonical ensemble, the charge is a
fixed parameter, and therefore the positivity of the heat capacity
or $(\partial T/\partial S)_{Q}=(\partial ^{2}M/\partial
S^{2})_{Q}$ is sufficient to ensure the local stability of the
system. In order to investigate the effects of the dilaton
coupling constant $\alpha$, the electric charge parameter $q$, and
the Born-Infeld parameter $\beta$ on the stability of the
solutions, we plot $(\partial ^{2}M/\partial S^{2})_{Q}$ versus
$\alpha$, $q$ and $\beta$. As one can see from figures
\ref{Figure8}-\ref{Figure11}, the topological black hole solutions
are stable independent of the values of the charge parameter $q$,
the curvature constant parameter $k$, and the Born-Infeld
parameter $\beta$, in any dimensions if $\alpha <\alpha_{\max }$,
while for $\alpha > \alpha_{\max }$ the system has an unstable
phase. That is the dilaton field makes the solution unstable.
These figures also show that $\alpha_{\max }$ increases with
increasing $q$ and $\beta$ (see Figs.
\ref{Figure8}-\ref{Figure10}), while it decreases with increasing
$k$ (see Fig. \ref{Figure11}). Again we kept fixed $l=b=1$ in
these figures. Note that as in the case of black hole solutions in
Einstein-Maxwell theory, the solutions are unstable for small
values of electric charge (see Figs. \ref{Figure12} and
\ref{Figure13}). That is the electric charge increases the stable
phase of the system. Besides, as the non-linearity of the
electromagnetic field increases ($\beta$ decreases), the stable
phase of the solution decreases. That is the non-linearity of the
electromagnetic field makes the solutions more unstable. This fact
may be seen in figures \ref{Figure14} and \ref{Figure15} which
show that the solutions are unstable for very highly nonlinear
field (small value of the Born-Infeld parameter).

\section{summary and conclusions}
The construction and analysis of topological black holes in AdS
space is a subject of much recent attention. This is primarily due
to their relevance for the AdS/CFT correspondence. In particular,
they allow us to study the dual conformal field theory on spaces
of the form $S^1 \times M^{d-2}$, where $M^{d-2}$ is an Einstein
space of positive, zero, or negative curvature \cite{Birm,Emp2}.
Apart the motivation comes from AdS/CFT side, the Born-Infeld
lagrangian coupled to a dilaton field appears very frequently in
string theory. Therefore it is of great importance to investigate
various properties of the topological black hole solutions in this
theory. In this paper, we constructed a new class of
$(n+1)$-dimensional charged topological black hole solutions of
Einstein-Born-Infeld-dilaton action. In contrast to the
topological black holes in the Einstein-Maxwell theory, which are
asymptotically AdS, the topological dilaton black holes we found
here, are neither asymptotically flat nor (A)dS. Indeed, the
Liouville-type potentials (the negative effective cosmological
constant) play a crucial role in the existence of these black hole
solutions, as the negative cosmological constant does in the
Einstein-Maxwell theory. In the $k=\pm1$ cases, these solutions do
not exist for the string case where $\alpha=1$. They are also
ill-defined for $\alpha =\sqrt{n}$. In the absence of a dilaton
field ($\alpha =0=\gamma $), our solutions reduce to the
$(n+1)$-dimensional topological Born-Infeld black hole solutions
presented in \cite{Cai1}, while in the limit $\beta\rightarrow
\infty$ they reduce to the topological black holes in
Einstein-Maxwell-dilaton gravity \cite{Sheykhi}. We computed the
conserved and thermodynamic quantities of the solutions and
verified that these quantities satisfy the first law of black hole
thermodynamics. Finally, we analyzed the thermal stability of
these black holes and disclosed the effect of the dilaton and
Born-Infeld fields on the stability of the solutions.

\acknowledgments{This work has been supported financially by
Research Institute for Astronomy and Astrophysics of Maragha.
Iran.}

\end{document}